\documentclass[reprint,cite,doublecol,figures,prb,amsmath,amssymb,superscriptaddress,showpacs]{revtex4-1}
\usepackage[dvips]{graphicx}
\DeclareGraphicsExtensions{.pdf} \graphicspath{{./figs/}}
\usepackage{color}
\usepackage{verbatim}
\usepackage[german,english]{babel}
\usepackage{pdfpages}

\usepackage{dcolumn}
\usepackage{bm}

\newcommand{\mr}{\mathrm}

\newcommand{\NTENPb}{Ni($N,N'$-bis(3-aminopropyl)propane-1,3-diamine)($\mu$-NO$_2$)ClO$_4$}

\newcommand{\NSM}{Neutron Scattering and Magnetism, Laboratory for Solid State Physics, ETH Z\"urich, CH-8093 Z\"urich, Switzerland}

\newcommand{\figref}[1]{Fig.\,\protect\ref{#1}}

\begin{document}


\title{Finite-temperature scaling of spin correlations in an experimental realization of the one-dimensional Ising quantum critical point}
\author{M.~H\"alg}
\thanks{Corresponding author}
\email{haelgma@phys.ethz.ch}
\affiliation{\NSM}
\author{D.~H\"uvonen}
\affiliation{\NSM}
\affiliation{National Institute of Chemical Physics and Biophysics, 12618 Tallinn, Estonia}
\author{T.~Guidi}
\affiliation{ISIS Facility, Rutherford Appleton Laboratory, Chilton, Didcot, Oxon OX11 0QX, United Kingdom}
\author{D.~L.~Quintero-Castro}
\affiliation{Helmholtz-Zentrum Berlin f\"ur Materialien und Energie, D-14109 Berlin, Germany}
\author{M.~Boehm}
\affiliation{Institut Laue-Langevin, BP 156, 38042 Grenoble, Cedex 9, France}
\author{L.~P.~Regnault}
\affiliation{Universit\'e Grenoble Alpes, CEA INAC-SPSMS-MDN, F-38000 Grenoble, France}
\author{M.~Hagiwara}
\affiliation{Center for Advanced High Magnetic Field Science, Graduate School of Science, Osaka University, 1-1 Machikaneyama, Toyonaka, Osaka 560-0043, Japan}
\author{A.~Zheludev}
\affiliation{\NSM}

\date{\today}

\begin{abstract}
Inelastic neutron scattering is used to study the finite-temperature scaling behavior of the local dynamic structure factor in the quasi-one-dimensional quantum antiferromagnet NTENP (\NTENPb), at its field-induced Ising quantum critical point.
The validity and the limitations of the theoretically predicted scaling relations are tested.

\end{abstract}

\maketitle

\section{\label{sec:Introduction}Introduction}
The Ising chain in a transverse field is one of the most-studied model systems for quantum phase transitions thanks to its simplicity and the possibility to obtain exact results.\cite{McCoy1968,Pfeuty1970,McCoy1983,Its1990,Sachdev1996,Sachdevbook}
It features a quantum critical point (QCP) at a certain value of magnetic field $H_c$ where the ground state changes from a quasi-classical ordered to a disordered quantum-paramagnetic phase.\cite{Sachdev1996,Sachdevbook}
Early theoretical studies examined the ground-state energy and spin correlations at zero temperature, $T=0$.\cite{McCoy1968,Pfeuty1970}
More recent work was able to derive explicit expressions for the finite-temperature correlation functions precisely at the critical field,\cite{McCoy1983,Its1990} and even in the quantum-paramagnetic and the magnetically ordered phases.\cite{Sachdev1996,Sachdevbook}
Independent of the details of the actual Hamiltonian, in the quantum critical regime, the spin correlations follow a universal scaling law.
Such universality is particularly interesting for experimentalists for two reasons.
On one hand, all predictions are expected to hold for \emph{any} material featuring a QCP in the universality class of Ising chains in a transverse field, irrespective of any material-dependent quantities.
On the other hand, spin correlations can be directly and accurately probed by neutron scattering experiments enabling an immediate comparison between experiment and theory.

Unfortunately, only few experimental realizations of spin-chain materials with strong uniaxial anisotropy and accessible field-induced QCPs are known.
Furthermore, most exhibit strong interchain coupling, as in the case of CoNb$_2$O$_6$\cite{Coldea2010} or Ca$_3$Co$_2$O$_6$\cite{Kageyama1997}.
Since strong interchain coupling leads to long-range order at finite temperatures already below $H_c$, these materials are only of limited use for studying the QCP.
Fortunately, the one-dimensional (1D) quantum Ising universality class -- which is equivalent to the two-dimensional (2D) \emph{thermodynamic} Ising universality class -- is also realized in the field-induced transition of Heisenberg $S=1$ chains with easy-plane anisotropy ($\mathcal{H}_{\text{aniso}} = \sum_i{D\left(S_i^z\right)^2}$, $D>0$) and with a magnetic field applied in an axial-symmetry-breaking way.\cite{Affleck1990,Affleck1991,Hieida2001,Sachdevbook}
An integer spin is necessary in combination with an easy-plane anisotropy in order to map the system to the one-dimensional quantum Ising model.
For this class of Hamiltonians, excellent experimental realizations are known, and have been studied in great detail.
The maybe best-known examples are Ni-based linear-chain azide complexes, e.g. NENP\cite{Regnault1994}, NMOAP\cite{Narumi1998}, NDMAP\cite{Honda1998,Chen2001} or NDMAZ\cite{Honda1997}.

\begin{figure}[!htb]
\unitlength1cm
\includegraphics[width=.3\textwidth]{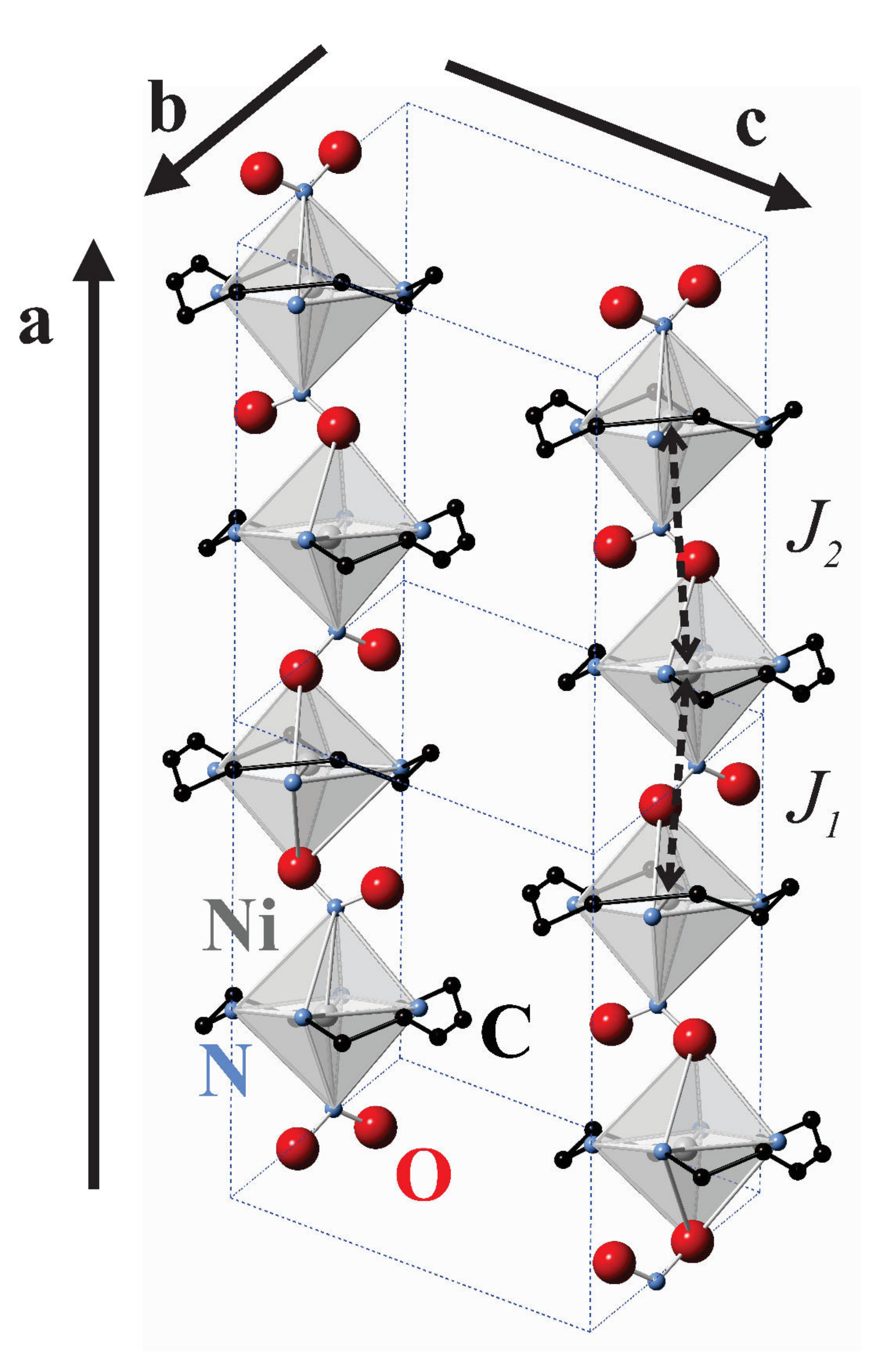}
\caption{Crystal structure of NTENP. The magnetic Ni$^{2+}$ ions are surrounded by octahedra with N$^{3-}$ ions in the equatorial vertices. This leads to a single-ion anisotropy with a hard axis along $a$. The chains run along $a$, where the Ni$^{2+}$ ions are linked by NO$_2^-$ groups. Perchlorate anions between the chains are omitted in the figure for clarity. Two unit cells are indicated by dotted lines. \label{fig:structure}}
\end{figure}

The objective of the present work is to experimentally test the finite-temperature scaling of the spin correlation function at the 1D Ising QCP with neutron scattering.
The target compound of this study is the antiferromagnetic (AF) $S=1$ chain material \NTENPb ~(NTENP).
The compound is well-characterized.
At room temperature, NTENP is a triclinic system with space group $P\bar{1}$, lattice constants $[a,b,c] = [10.747(3),9.413(2),8.789(2)]~\mr{\AA}$ and angles $[\alpha,\beta,\gamma] = [95.52(2)^{\circ},108.98(3)^{\circ},106.83(3)^{\circ}]$.\cite{Escuer1997}
It features bond-alternating AF exchange with exchange constants $J_1 = 2.1~\text{meV}$ and $J_2 = 4.7~\text{meV}$ as well as single-ion anisotropy, $D = 1.2~\text{meV}$.\cite{Narumi2001,Zheludev2004,Regnault2004,Regnault2006}
Interchain interaction are only weak, being of the order of $0.005 J_1$.\cite{Zheludev2004}
The magnetic properties of NTENP can be described by the following leading terms in the Hamiltonian
\begin{equation}
\mathcal{H} = \sum_{i=1}^{N/2}\left(J_1 \textbf{S}_{2i-1} \textbf{S}_{2i} + J_2 \textbf{S}_{2i} \textbf{S}_{2i+1}\right) +
\sum_{i=1}^ND\left(S_i^z\right)^2.
\end{equation}
The chains along the crystallographic $a$ axis are formed by Ni$^{2+}$ ions which interact via NO$_2^-$ groups.\cite{Escuer1997}
The crystal structure is shown in \figref{fig:structure}.
The material has a spin-dimerized non-magnetic ground state and the excitation spectrum is, accordingly, gapped.
No magnetic order can be observed down to 50~mK in the absence of a magnetic field.
The first excited states are an $S=1$ triplet, split by single-ion anisotropy into a lower-lying doublet with a gap of 1.06~meV and a higher-energy singlet at 1.96~meV.\cite{Hagiwara2005,Regnault2006}
Applying an external magnetic field decreases the energy gap in one member of the doublet due to the Zeeman effect.
For an axial-symmetry-preserving magnetic field applied along the hard axis, i.e. perpendicular to the easy plane, a transition to the gapless Tomonaga-Luttinger spin liquid phase occurs at a critical magnetic field of 93~kOe.\cite{Hagiwara2006}
However, previous experiments showed that if the magnetic field is applied in the easy plane, NTENP undergoes an Ising quantum phase transition at $H_c \approx 113~\text{kOe}$.\cite{Hagiwara2005,Regnault2006}
The high-field phase is gapped and antiferromagnetically ordered.
Due to its experimentally accessible critical field and small interchain interactions, NTENP in a magnetic field applied in the easy plane appears to be a good candidate for the present study.

In this paper, the focus is on the local or $q$-integrated dynamic structure factor,
\begin{equation}
S\left(\omega\right) = \int_{-\infty}^{\infty}{\left\langle S_i^z\left(t\right),S_i^z\left(0\right) \right\rangle e^{i\omega t}dt},
\end{equation}
which we measure in the temperature range between 0.027~K and 12.5~K, for neutron energy transfers $\hbar\omega$ up to 1.7~meV.
These experimental results are used to test the predicted scaling law, and to probe the limits of its applicability.

\section{Experimental details}
Half-deuterated (i.e. about 50~\% of the hydrogen atoms replaced by deuterium) single-crystal samples of NTENP, from the same batch as the samples used for the experiments of reference~\cite{Zheludev2004}, were characterized by single-crystal x-ray diffraction.
The data were collected on a Bruker AXS single-crystal x-ray diffractometer employing a Cryostream 700 from Oxford Cryosystems for temperature control between 300~K and 100~K.
The samples were additionally studied in calorimetry experiments.
We employed a commercial Quantum Design physical property measurement system (PPMS) for collecting specific-heat data in the range between 300~K and 1.8~K in zero magnetic field.
Contributions to specific heat from the setup and the grease used for mounting the sample were subtracted from the data.
The temperature rise for each data point was 2~$\%$ of the sample temperature.

Neutron scattering experiments were performed on a fully-deuterated single crystal of mass 0.5~g which was previously used for the experiments reported in reference~\cite{Regnault2006}.
The crystal was aligned with the neutron beam perpendicular to the crystallographic $b$ axis.
Neutron data were collected on the three-axes spectrometers IN14 at ILL and FLEXX at HZB,\cite{Le2013} as well as on the time-of-flight multi-chopper spectrometer LET at ISIS.
A 150~kOe (IN14), a 148~kOe (FLEXX) and a 140~kOe (LET) vertical magnet ($\textbf{H}~||~\textbf{b}$) with dilution insert were used.
The final neutron momentum $k_f$ was fixed to 1.3~$\text{\AA}^{-1}$ in the experiments on IN14 and FLEXX.
On LET the incident neutron energy was set to 2.2~meV.
With these setups, the full width at half maximum (FWHM) of the quasi-elastic line of NTENP was measured to be 0.12~meV (IN14), 0.10~meV (FLEXX) and 0.05~meV (LET), respectively.
For each experiment, the critical magnetic field $H_c$ was determined independently by following the field dependence of the magnetic Bragg peak at $q=(1,0,0)$ in the ordered phase at $H>H_c$.

\section{Experimental results}
\begin{figure}[!htb]
\unitlength1cm
\includegraphics[width=.48\textwidth]{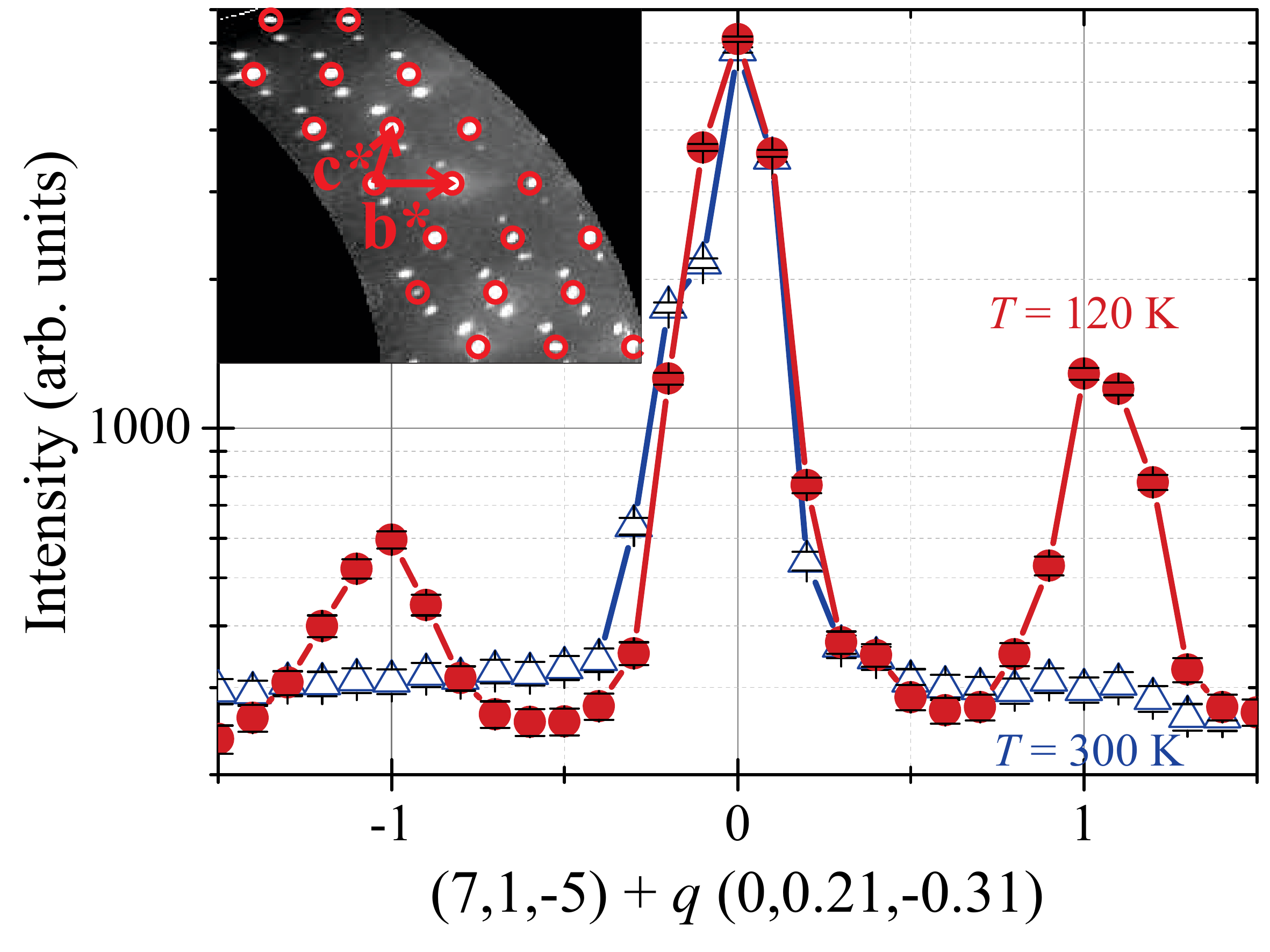}
\caption{X-ray diffraction data measured in NTENP in the vicinity of the $(7,1,-5)$ Bragg peak at 300~K and 120~K. Two satellite peaks are clearly visible at 120~K. The inset shows the $7kl$-plane at 120 K where red circles mark the principle Bragg peaks of the room-temperature crystal structure. \label{fig:peaks}}
\end{figure}

\begin{figure}[!htb]
\unitlength1cm
\includegraphics[width=.48\textwidth]{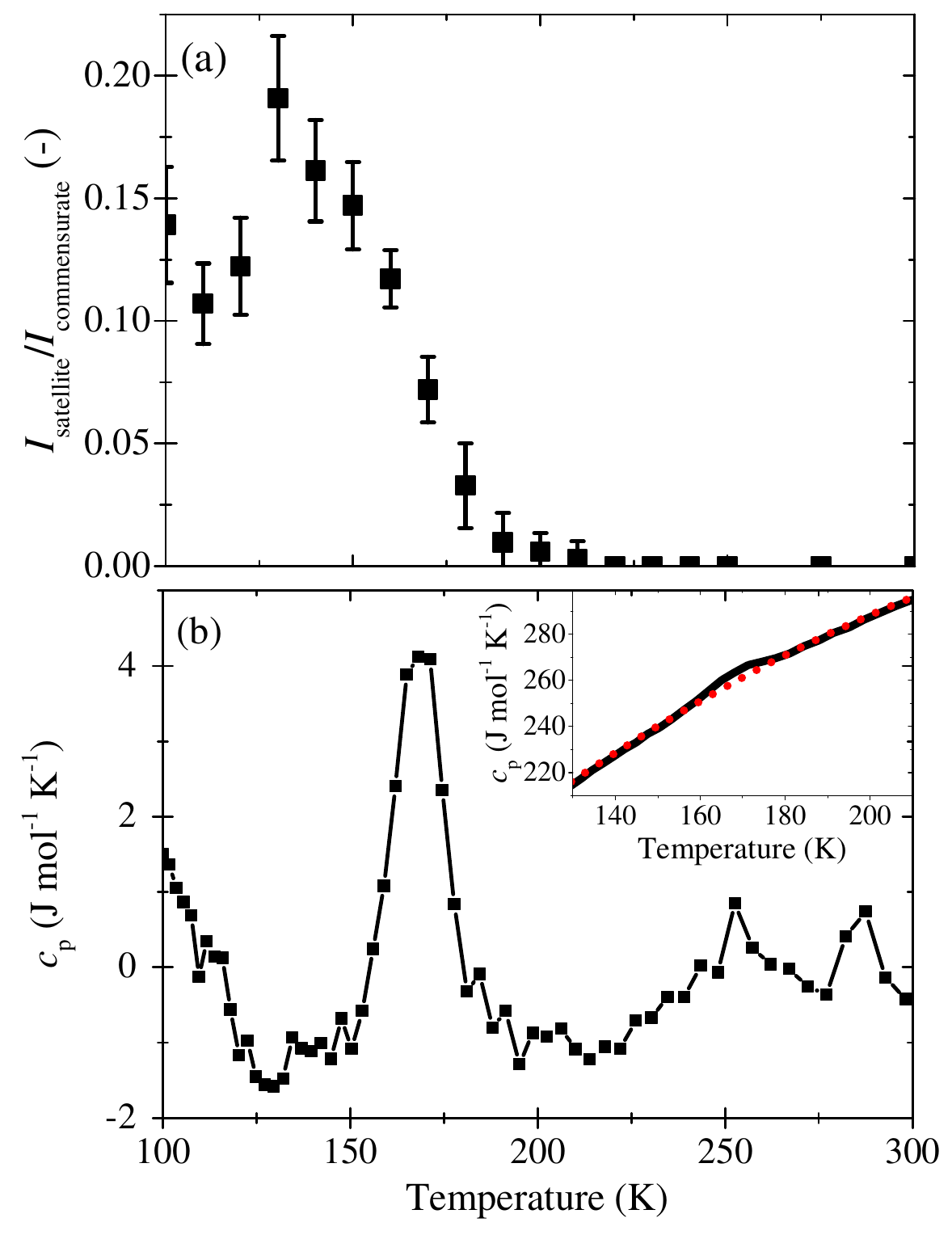}
\caption{(a) The temperature dependence of the satellite-peak intensity measured in NTENP. (b) The specific heat features a maximum at about 170~K. To emphasize the maximum visually, a second-order polynomial function was subtracted from the total specific heat. The inset illustrates the total specific heat (black, solid line) and the fit (red, dotted line).
\label{fig:transition}}
\end{figure}

X-ray measurements on half-deuterated single crystals revealed new incommensurate structural satellite peaks appearing in the $kl$-plane for all integer $h$ at low temperatures (\figref{fig:peaks}).
The incommensurate peaks persist at least down to 0.5~K as was observed in the neutron scattering experiment on LET.
They reside at $\delta q = \pm \left[0.00(2),0.21(3),-0.31(2)\right]$ around the commensurate nuclear Bragg peaks and appear upon cooling at about 180~K as is illustrated in \figref{fig:transition}(a), signaling a previously unknown structural transition.
The occurrence of a phase transition is further confirmed by specific-heat measurements (\figref{fig:transition}(b)).
The exact values of the transition temperature and of the satellite-peak intensity is not fully reproducible.
Moreover, the non-monotonous behavior of the satellite-peak intensity is not understood.
Nevertheless, below 180~K the superstructure is present in all samples studied.
Our data is clearly insufficient to understand the details of the phase transition and the newly established structure.
This issue may be subject to a further study.
However, for the present study the precise nature of the phase transition is not important, since we only focus on \emph{universal} magnetic properties at the QCP.
As will be discussed below, for us, the only relevant consequence of the transition is the residual spin gap at $H_c$.

\begin{figure}[!htb]
\unitlength1cm
\includegraphics[width=.48\textwidth]{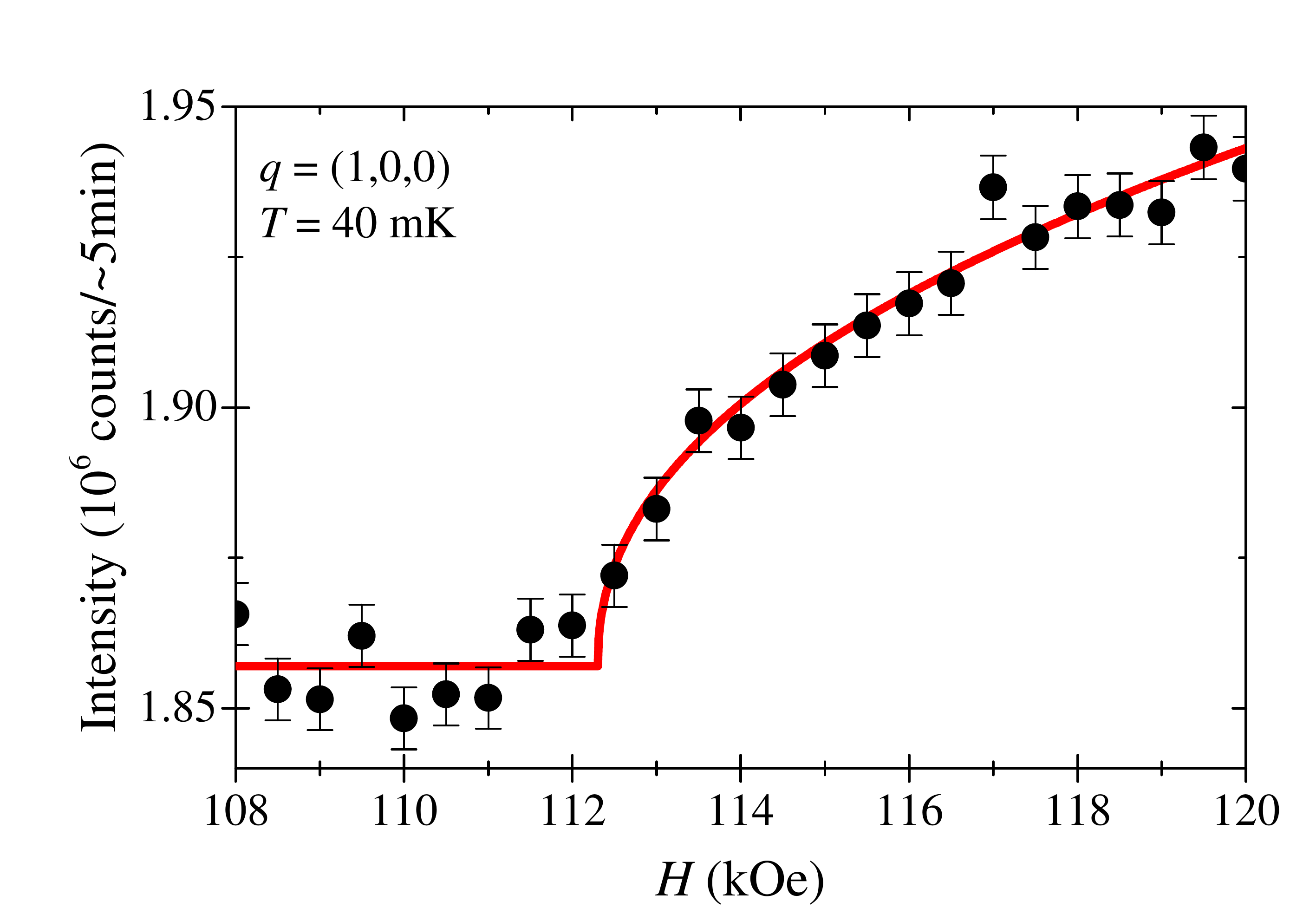}
\caption{Peak intensity of the $(1,0,0)$ Bragg reflection at 40~mK versus magnetic field, as measured on FLEXX / HZB. Above a critical magnetic field $H_c|_{T=40~\text{mK}}$ the system is magnetically ordered and exhibits a magnetic Bragg peak at $q=(1,0,0)$. The solid line is an empirical power-law fit to the data with fitted exponent 0.5(2) yielding a critical magnetic field of $H_c|_{T=40~\text{mK}} =$~112.3(5)~kOe~$>H_c$. \label{fig:Hc}}
\end{figure}

The main focus of the present work is on the dynamic structure factor at the QCP.
A crucial point is an accurate determination of the critical field $H_c$.
It can be determined by observing the magnetic Bragg peaks due to AF order stabilized in the high-field phase by weak interchain interactions.\cite{Hagiwara2005}
Since there are two spins per lattice period along the $a$ axis (see \figref{fig:structure}), the AF Bragg peaks occur at integer values of $h$.
To determine $H_c$, at the start of each neutron experiment, the intensity of the $q=(1,0,0)$ Bragg reflection was followed versus magnetic field at a temperature close to the base temperature.
Exemplarily, data collected on FLEXX are shown in \figref{fig:Hc}.
The critical field was estimated using an empirical power-law fit (solid line in \figref{fig:Hc}).
The actual zero-temperature critical field has to be slightly smaller than the magnetic field where the magnetic Bragg peak appears.
Nevertheless, since the system remains in the quantum critical regime as long as the temperature is large enough compared to the energy gap, small deviations of the used magnetic field from $H_c$ are acceptable.
The magnetic field chosen for the experiments are 112~kOe (FLEXX) and 113~kOe (IN14 and LET) respectively.

\begin{figure}[!htb]
\unitlength1cm
\includegraphics[width=.48\textwidth]{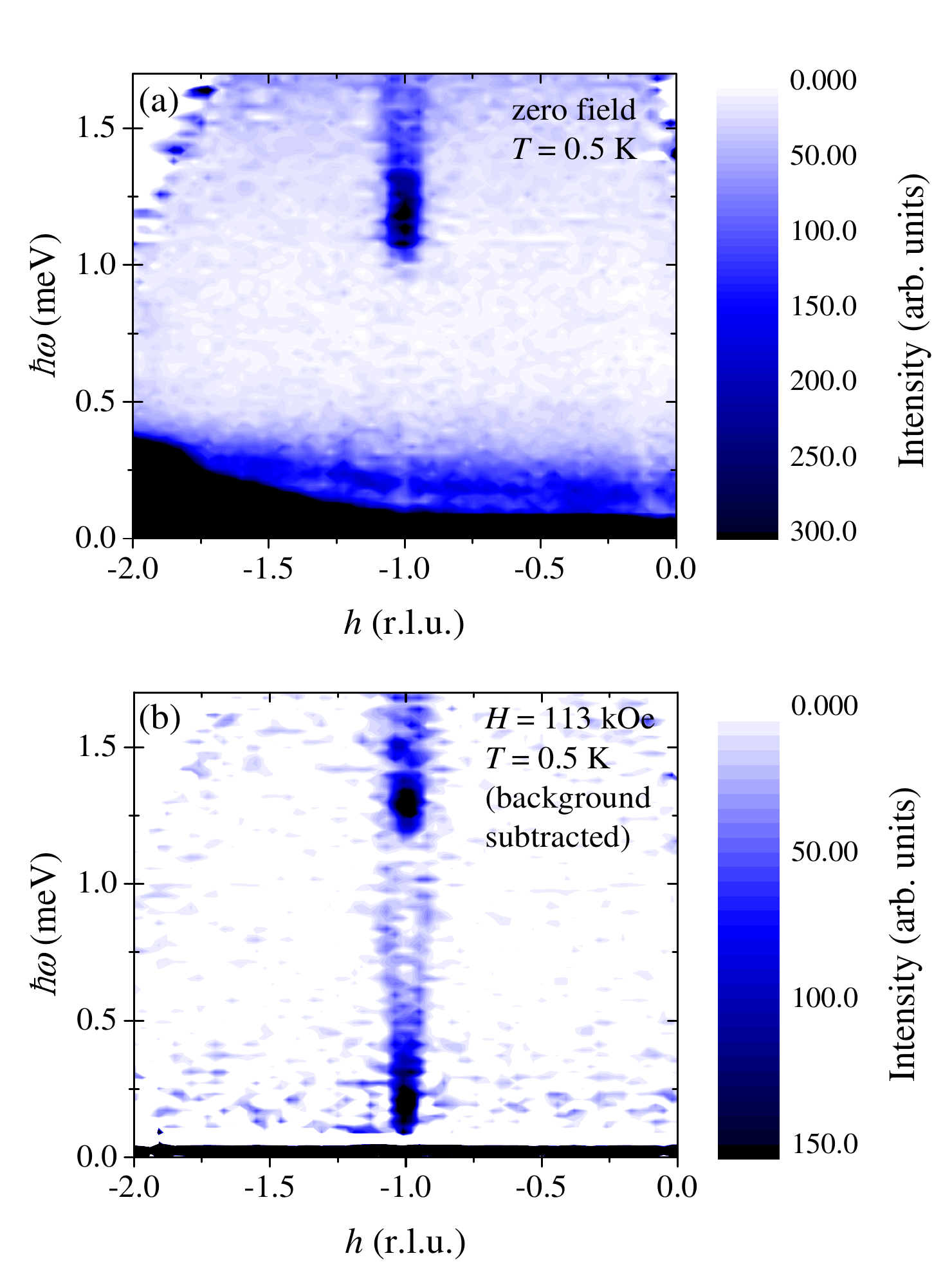}
\caption{The low-energy spectra of NTENP at 0.5~K measured on LET / ISIS are shown in zero field (a) and at 113~kOe (b). The zero-field spectrum was used as the background for the measurements in field. In magnetic field, the $S=1$ triplet states are split due to the Zeeman effect and the excitation energy of the lowest excited state is reduced whereas one mode persists at $\Delta_0 \approx$~1.2~meV. The residual energy gap which can be observed at 113~kOe is discussed in section~\ref{sec:gap}. \label{fig:spectrum}}
\end{figure}

\begin{figure}[!htb]
\unitlength1cm
\includegraphics[width=.48\textwidth]{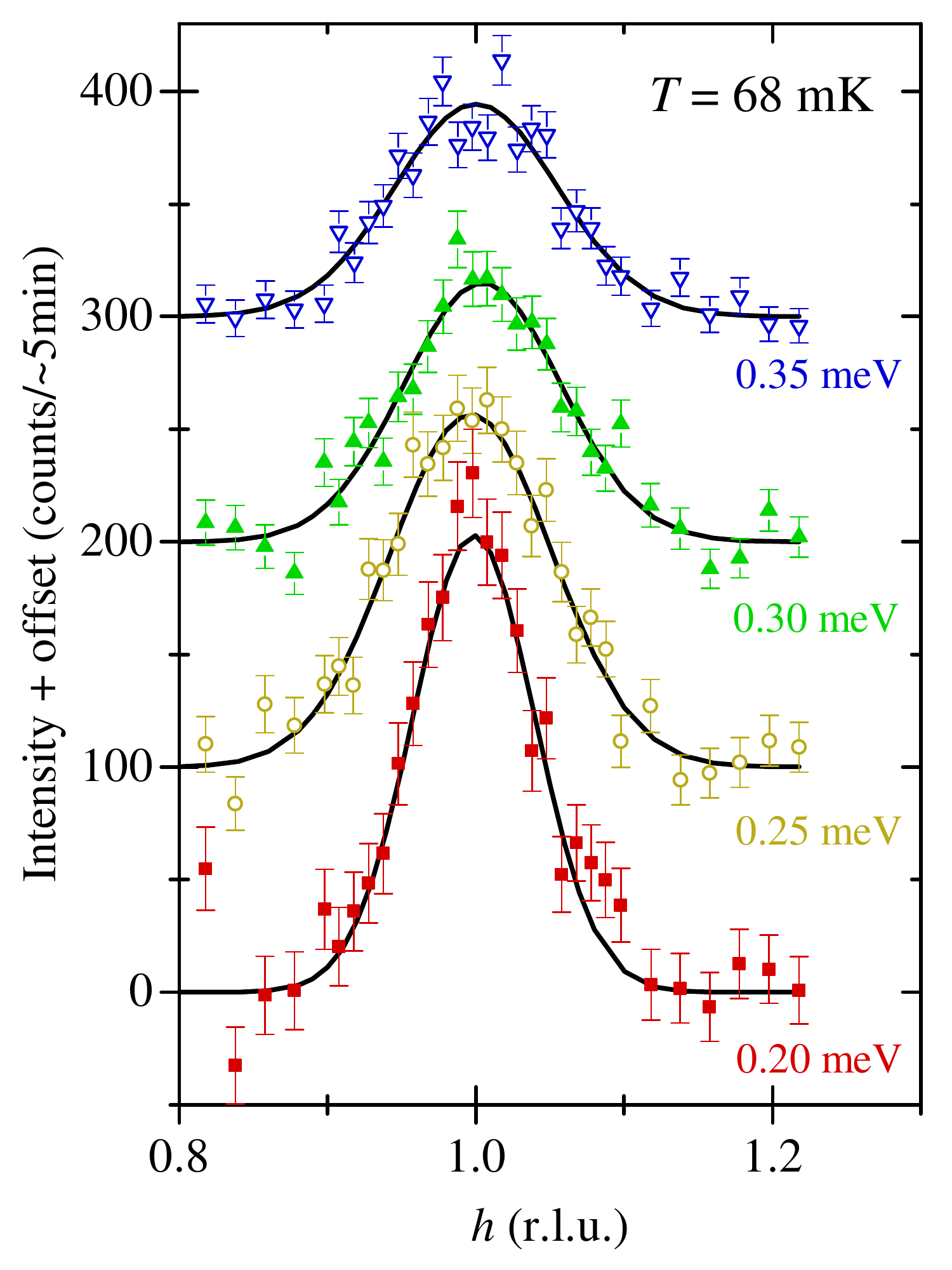}
\caption{Background-subtracted constant-energy data measured in NTENP at $H = 112~\text{kOe} \approx H_c$ on FLEXX / HZB at 68~mK around the AF wave vector $h=1$. A constant offset of 100~counts/$\sim$5min is applied between data measured at different energy transfers. Solid, black lines are guides to the eye.\label{fig:dataV2}}
\end{figure}

\begin{figure}[!htb]
\unitlength1cm
\includegraphics[width=.48\textwidth]{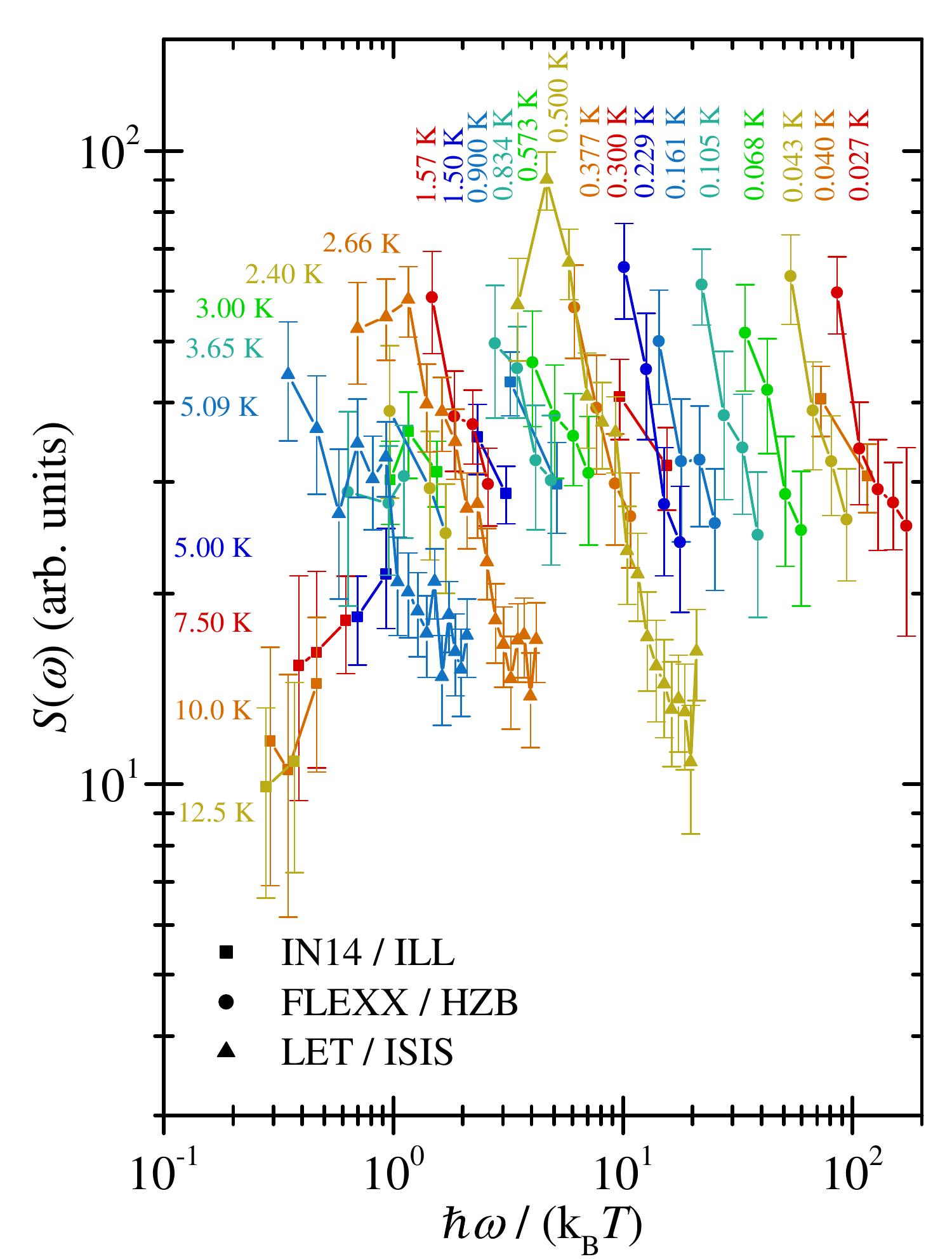}
\caption{Summary of the neutron-scattering data for NTENP at the critical magnetic field obtained on IN14, FLEXX and LET. The $q$-integrated dynamic structure factor is shown versus the temperature-scaled energy transfer $\hbar\omega/\left(\text{k}_{\text{B}}T\right)$. \label{fig:unscaled}}
\end{figure}

On the time-of-flight spectrometer LET the full low-energy spectrum could be obtained in the vicinity of the AF wave vector.
It was collected at 113~kOe at 0.50~K, 2.66~K and 5.09~K.
Additionally, the spectrum in zero field was measured at base temperature (0.5~K) and used as the background for the measurements in field.
Respective data are shown in \figref{fig:spectrum}.
On the three-axes spectrometers IN14 and FLEXX constant-energy scans were performed in magnetic field at energies between 0.15~meV and 0.40~meV for various temperatures in the range between 0.027~K and 12.5~K.
On FLEXX, the background was measured at 0.15~K in zero field and subtracted from the raw data (\figref{fig:dataV2}).
In contrast, the background for the measurements on IN14 was extracted from the data obtained in field.
For this purpose, for each energy a polynomial function was fitted to the data at all different temperatures simultaneously, not taking into account the excitation in the window $0.85 \leq h \leq 1.15$.
The data was integrated numerically along $h$ in order to obtain the $q$-integrated or local dynamic structure factor $S\left(\omega\right)$.
Figure~\ref{fig:unscaled} summarizes all data for $S\left(\omega\right)$ collected on the three spectrometers for all measured temperatures and energies $\hbar\omega < 0.9~\text{meV}$.

\section{Discussion\label{sec:discussion}}

\subsection{\label{sec:gap}Residual energy gap at $H_c$}
The spectrum at 113~kOe measured on LET (\figref{fig:spectrum}(b)) clearly exhibits an energy gap, $\Delta \approx$~0.2~meV, although the system is expected to be gapless.
A possible explanation of the appearance of the gap is that the applied magnetic field deviates from the actual critical field.
However, no closing of the gap was observed in several measurements between 105~kOe and 120~kOe (not shown).
Therefore, we conclude that the gap does not close at any value of applied magnetic field at all.
The fact that no residual gap has been observed in previous studies\cite{Hagiwara2005,Regnault2006} is most certainly due to an insufficient energy resolution in the chosen configurations of those experiments performed on three-axes spectrometers.

The most probable cause for the residual energy gap close to the critical field is the observed structural phase transition reported in this work.
One possible effect is the formation of additional anisotropies like Dzyaloshinskii-Moriya (DM) interactions, an anisotropic $g$-tensor or in-plane anisotropies evoking a staggered magnetic field which prevents the gap from closing like in the case of the related compound NENP.\cite{Chiba1991,Mitra1994}
Indeed, nuclear magnetic resonance (NMR) experiments on NTENP revealed a field-induced inhomogeneous magnetization for a magnetic field applied in the $ab^{\ast}$-plane.\cite{Matsubara2005}

\subsection{The scaling law and its empirical range of applicability}
Close to a QCP the $q$-integrated dynamic structure factor should not depend on any properties of the system under investigation but is determined by the underlying universality class.
The only energy scale, for example, is provided by the temperature $T$ itself.
As a result, it can be written in a scaling form:
\begin{equation}
S\left(\omega\right) = T^{-\alpha}\Phi\left(\frac{\omega}{T}\right),
\label{eq:scalinglaw}
\end{equation}
where the scaling function $\Phi$ and the scaling exponent $\alpha$ are specific to the universality class.\cite{Sachdevbook}
This formula implies that all $q$-integrated neutron data measured at different temperatures should collapse on a single curve given by $\Phi$ when plotted versus $\omega/T$ and scaled by a factor $T^{\alpha}$.

In practice, one has to exercise caution.
The residual energy gap implies that the system is never at the QCP in the experiments of the present work.
Despite that, the properties of the QCP can still be probed at temperatures or energy transfers (frequencies) that are large compared to the gap energy such that the system is in the quantum critical regime, i.e.
\begin{equation}
\left(\text{k}_{\text{B}}T \gg \Delta\right) \text{ or } \left(\hbar\omega \gg \Delta\right).
\label{eq:scaling1}
\end{equation}
An additional upper limit for temperature and energy is due to the presence of the non-critical spin fluctuations at higher energy transfers, $\Delta_0 \approx$~1.2~meV, as shown in \figref{fig:spectrum}(b).
At too large temperatures and energy transfers, this higher mode is excited and the corresponding scattering cannot be separated from the critical part.
Thus, the second scaling condition to be fulfilled is
\begin{equation}
\text{k}_{\text{B}}T,\hbar\omega \ll \Delta_0.
\label{eq:scaling2}
\end{equation}
Only data fulfilling the scaling conditions, equations (\ref{eq:scaling1}) and (\ref{eq:scaling2}), can be considered for testing the scaling relations.
The practical upper and lower boundaries were determined empirically to be
\begin{equation}
\left(T > \text{2.9~K}\right) \text{ or } \left(\hbar\omega > \text{0.25~meV}\right),
\label{eq:scaling1emp}
\end{equation}
\begin{equation}
\left(T < \text{7.5~K}\right) \text{ and } \left(\hbar\omega < \text{0.9~meV}\right).
\label{eq:scaling2emp}
\end{equation}

\begin{figure}[!htb]
\unitlength1cm
\includegraphics[width=.48\textwidth]{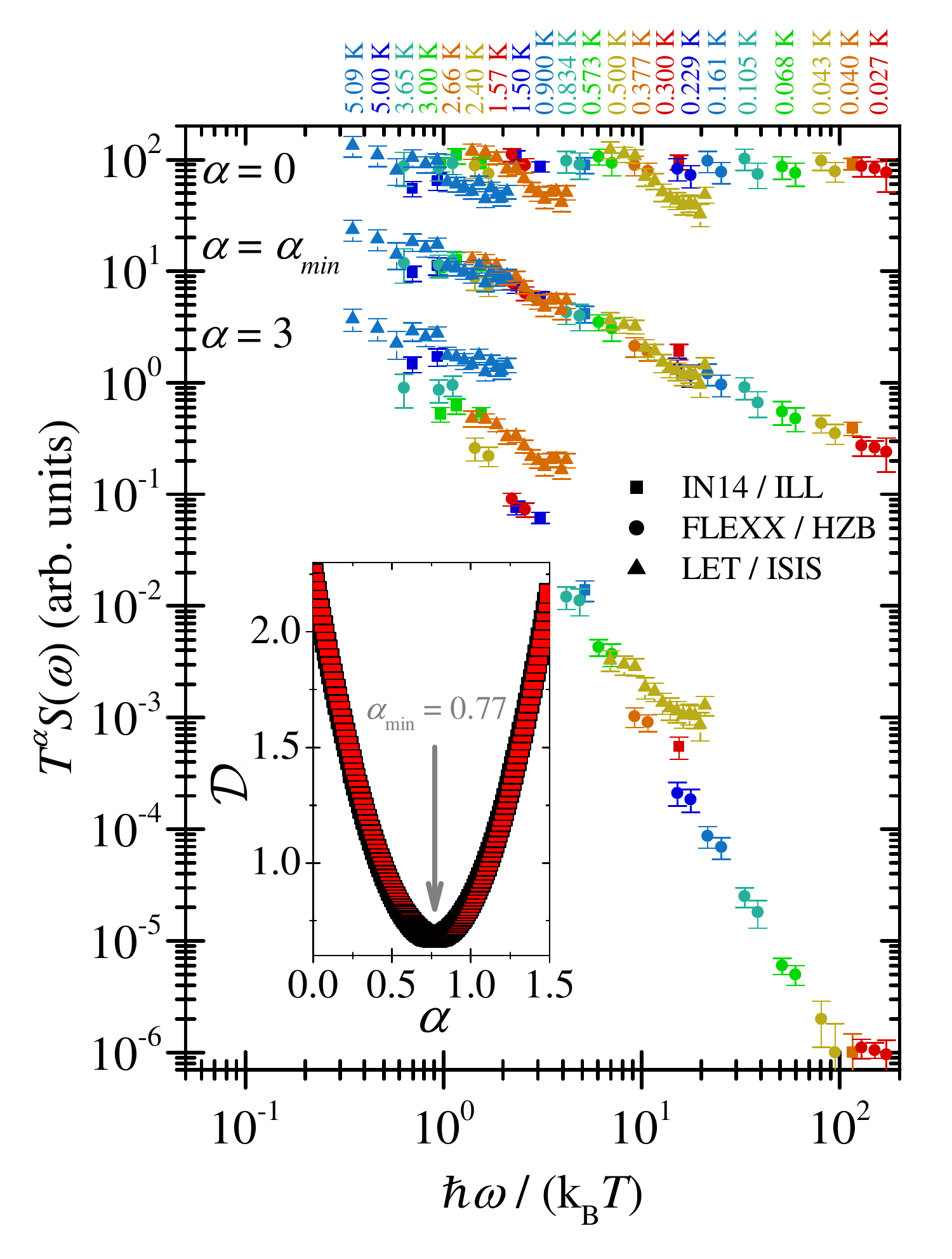}
\caption{Scaled $q$-integrated dynamic structure factor $S\left(\omega\right)$ measured in NTENP and plotted with scaling exponents $\alpha=0$, 3 and $\alpha_{\text{min}}$. The scaling exponent with the best data overlap, $\alpha_{\text{min}}=0.77(2)$, was determined by minimizing the function $\mathcal{D}\left(\alpha\right)$ (inset).\cite{supplemental} For the data with $\alpha=0$ an additional constant prefactor is applied in order to better discern the data with $\alpha=0$ and $\alpha_{\text{min}}$. \label{fig:scalingfit}}
\end{figure}

In order to determine the scaling exponent $\alpha$, the quality of the overlap between data sets collected at different temperatures has to be quantified.
For this purpose, an approach similar to the one in reference~\cite{Povarov2015} was used.
As described in detail in the supplementary material,\cite{supplemental} a $\chi^2$-like measure $\mathcal{D}\left(\alpha\right)$ is introduced for the data mismatch in the scaling plot produced with a particular scaling exponent $\alpha$.
Figure~\ref{fig:scalingfit} shows the scaled data for three different scaling exponents $\alpha$ as well as the mismatch function $\mathcal{D}\left(\alpha\right)$ for $\alpha$ in the range between 0 and 1.5.
It is minimal at $\alpha_{\text{min}}=0.77(2)$ implying that the best data overlap is achieved for this value of the scaling exponent.

\subsection{1D Ising QCP}
\begin{figure}[!htb]
\unitlength1cm
\includegraphics[width=.48\textwidth]{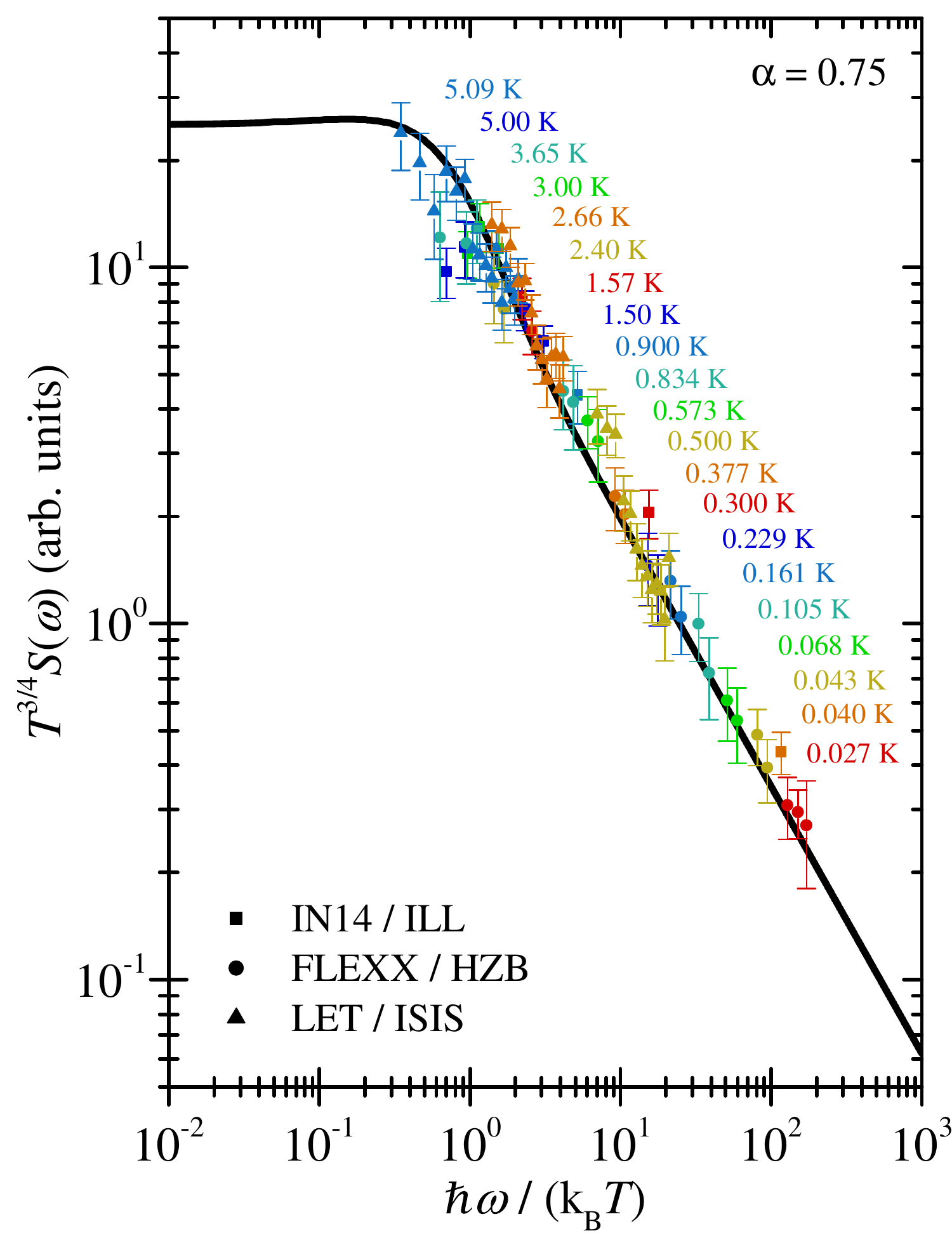}
\caption{The scaling plot for the local dynamic structure factor measured in NTENP with the Ising scaling exponent $\alpha_{\text{Ising}} = 0.75$ is depicted. The black line corresponds to the predicted universal scaling function for the quantum critical point of an Ising chain in a transverse magnetic field as described in the text, Eq.~(\ref{eq:phi}). \label{fig:scaling}}
\end{figure}

The observed scaling form for the local spin correlations at the QCP in NTENP can now be compared to specific predictions for the 1D Ising QCP.
For this model, the transverse dynamic structure factor is\cite{Sachdev1996,Sachdevbook}
\begin{equation}
\begin{split}
S\left(\tilde{q},\omega\right) \propto~ & T^{-\frac{7}{4}} \left(\frac{1}{e^{\frac{\hbar\omega}{\text{k}_{\text{B}}T}}-1}+1\right) \times\\
& \Im\left(\frac{\Gamma\left(\frac{1}{16} - i\frac{\hbar\omega+u\tilde{q}}{4\pi \text{k}_{\text{B}}T}\right) \Gamma\left( \frac{1}{16} - i\frac{\hbar\omega-u\tilde{q}}{4\pi \text{k}_{\text{B}}T} \right)}{\Gamma\left(\frac{15}{16} - i\frac{\hbar\omega+u\tilde{q}}{4\pi \text{k}_{\text{B}}T}\right) \Gamma\left( \frac{15}{16} - i\frac{\hbar\omega-u\tilde{q}}{4\pi \text{k}_{\text{B}}T} \right)}\right),
\end{split}
\label{eq:chi}
\end{equation}
where $\tilde{q}$ is the momentum transfer relative to the critical wave vector, $u$ is the spin wave velocity, $\Gamma$ is the gamma function and $\Im$ denotes the imaginary part of the argument.
Integrating over $q$ yields the scaling form,
\begin{equation}
S\left(\omega\right) = T^{-\frac{3}{4}} \Phi\left(\frac{\omega}{T}\right),
\label{eq:Sw}
\end{equation}
with the critical exponent $\alpha_{\text{Ising}} = 0.75$ and the scaling function given by
\begin{equation}
\begin{split}
\Phi\left(\frac{\omega}{T}\right) \propto	&	\left(\frac{1}{e^{\frac{\hbar\omega}{\text{k}_{\text{B}}T}}-1}+1\right) \times\\
	&	\Im\left( \frac{\Gamma\left(\frac{1}{8}-i\frac{\hbar\omega}{2\pi \text{k}_{\text{B}}T}\right)}{\Gamma\left(\frac{7}{8}\right)^2\Gamma\left(\frac{7}{8}-i\frac{\hbar\omega}{2\pi \text{k}_{\text{B}}T}\right)} \right).
\end{split}
\label{eq:phi}
\end{equation}
The experimentally identified scaling exponent for NTENP, $\alpha = 0.77(2)$, agrees remarkably well with this expectation.
Figure~\ref{fig:scaling} shows the scaling plot for NTENP with the Ising scaling exponent $\alpha_{\text{Ising}}=0.75$.
Additionally, the solid line represents a fit of the scaling function given in Eq.~(\ref{eq:phi}) with an overall prefactor (vertical shift in the $\log$-$\log$ plot) as the only adjustable parameter.
Once again, an excellent quantitative agreement with the data is apparent.

\section{Conclusions}
In summary, the scaling relation for the local dynamic structure factor of a linear-chain antiferromagnet has been verified at a field-induced Ising QCP over three orders of magnitude in $\hbar\omega/\left(\text{k}_{\text{B}}T\right)$.
The use of neutron spectroscopy and the coverage of a wide range of energy transfers allows to evade non-critical spin fluctuations.
It also provides a way to avoid the unwanted terms in the Hamiltonian that prevent an exact realization of the 1D Ising QCP in any realistic material.
In this respect, neutron spectroscopy shows more flexibility than other techniques, e.g. NMR, which essentially measures the same local dynamic structure factor but is restricted to a single, low measurement frequency.\cite{Mukhopadhyay2012,Kinross2014}

\section{Acknowledgments}
This project was supported by division II of the Swiss National Science Foundation.
It has received funding from the European Union's Seventh Framework Program for research, technological development and demonstration under the NMI3-II Grant number 283883.
Parts of this work are based on experiments performed at the Institut Laue-Langevin (ILL) in Grenoble, France, the Helmholtz Zentrum Berlin (HZB), Germany, and ISIS at the STFC Rutherford Appleton Laboratory in Oxfordshire, UK.
Additionally, this work was partly supported by the Estonian Ministry of Education and Research under grant No. IUT23-03, the Estonian Research Council grant No. PUT451 and the Grants-in-Aid for Scientific Research (Nos. 242440590 and 25246006) from the MEXT, Japan.
Finally, we would like to thank F. Essler for valuable discussions and K. Yu. Povarov for his help in the dielectric characterization of the compound.


\centerline{***}

\bibliography{bibliography}\label{Cbib}

\onecolumngrid
\newpage

\includepdf{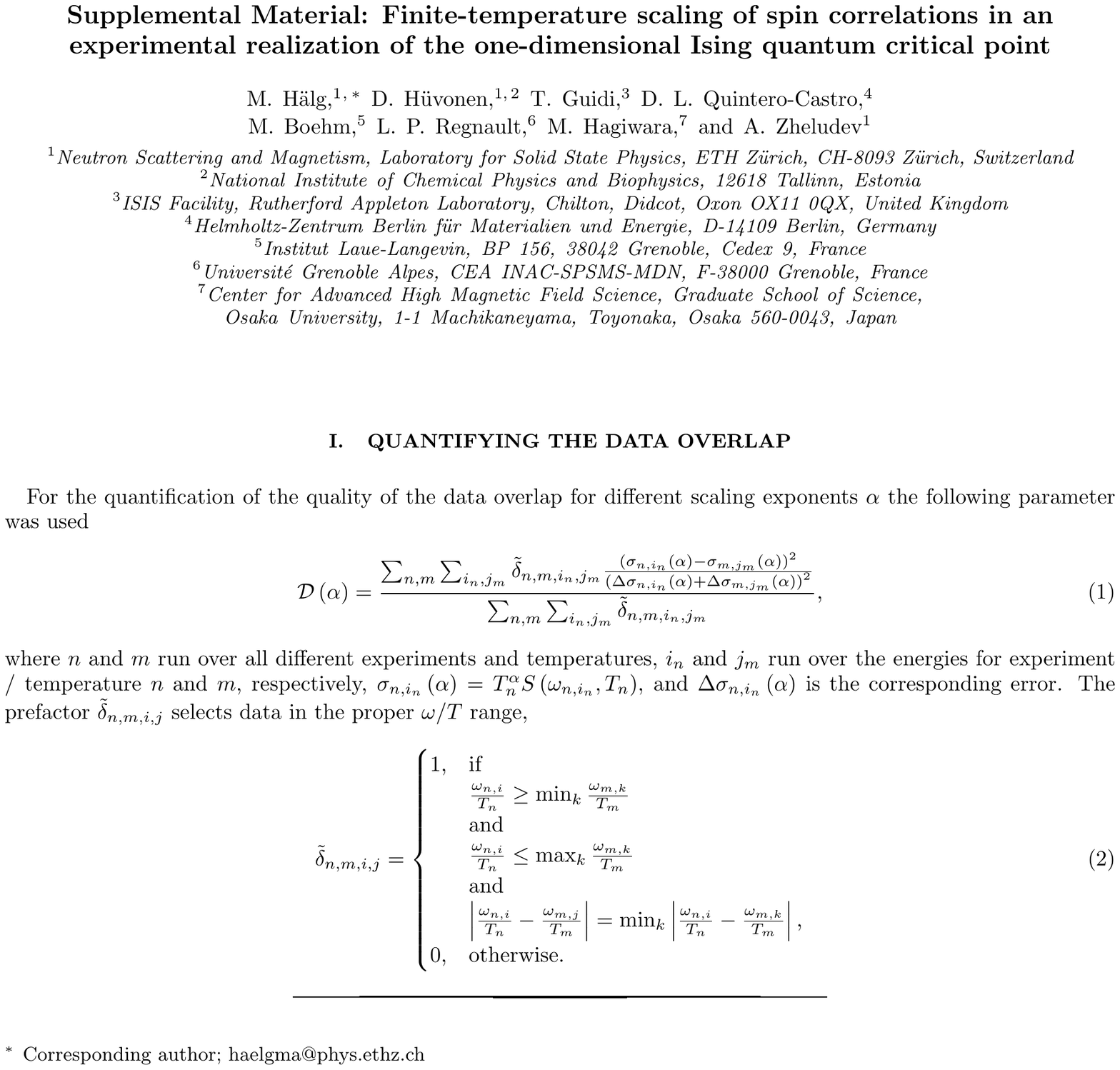}

\end{document}